\documentclass[prl,twocolumn,showpacs,superscriptaddress,floatfix]{revtex4}
\usepackage{graphicx,amsfonts,amssymb,amsmath,times}
\usepackage[colorlinks,bookmarks=false,citecolor=blue,linkcolor=red,urlcolor=blue]{hyperref}
\usepackage{color}

\newlength{\ldag}
\begin{document}

\title{Effective spin model for the spin-liquid phase of the Hubbard model on the triangular lattice}

\author{Hong-Yu Yang}
%\email{yang@fkt.physik.tu-dortmund.de}
\affiliation{Lehrstuhl f\"ur Theoretische Physik I, Otto-Hahn-Stra\ss e 4, TU Dortmund, D-44221 Dortmund, Germany}

\author{Andreas M. L\"auchli}
%\email{laeuchli@comp-phys.org}
\affiliation{Max Planck Institut f\"ur Physik komplexer Systeme, D-01187 Dresden, Germany}

\author{Fr\'ed\'eric Mila}
%\email{frederic.mila@epfl.ch}
\affiliation{Institute of Theoretical Physics, Ecole Polytechnique F\'ed\'erale de Lausanne, CH-1004 Lausanne, Switzerland}

\author{Kai Phillip Schmidt}
%\email{schmidt@fkt.physik.tu-dortmund.de}
\affiliation{Lehrstuhl f\"ur Theoretische Physik I, Otto-Hahn-Stra\ss e 4, TU Dortmund, D-44221 Dortmund, Germany}

%------------------------------------------------------------------------------

\begin{abstract}
We show that the spin liquid phase of the half-filled Hubbard model on the triangular lattice
can be described by a pure spin model. This is based on a high-order strong coupling expansion
(up to order 12) using perturbative continuous unitary transformations. The resulting spin model is consistent
with a transition from three-sublattice long-range magnetic order to an insulating spin liquid phase,
and with a jump of the double occupancy at the transition. Exact diagonalizations of
both models show that the effective spin model is quantitatively accurate well into the spin liquid phase,
and a comparison with the Gutzwiller projected Fermi sea suggests a gapless spectrum and a spinon Fermi surface.
\end{abstract}


\maketitle
%
%
%%%%%%%%%%%%%%
%\emph{Introduction ---}
%%%%%%%%%%%%%%
Although the Hubbard model has been one of the central paradigms in the field of strongly correlated systems
for about five decades, new aspects of its extremely rich phase diagram are regularly unveiled. Even at
half-filling, the popular wisdom according to which the model has only two phases, a metallic one at weak
coupling and an insulating one at strong coupling separated by a first order transition \cite{Georges96}, has been
recently challenged. This goes back to the work of Morita {\it et al.} \cite{Morita02} on the
triangular lattice which revealed the presence of a non-magnetic insulating phase close to the metal-insulator
transition using path integral renormalization group. Further evidence in favour of a phase transition inside the insulating phase has been reported using a variety of theoretical tools \cite{Kyung06,Sahebsara08,Tocchio08,Yoshioka09}. More recently, an intermediate spin liquid (SL) phase has also been identified on the honeycomb lattice using Quantum Monte Carlo simulations \cite{Meng10}.

The precise nature of the SL phase of the Hubbard model on the triangular lattice is of direct experimental
relevance for the 2D organic salt $\kappa$-(BEDT-TTF)$_{2}$Cu$_{2}$(CN)$_{3}$ \cite{Shimizu03}.
As such, it has already attracted a lot of attention, but fundamental questions such as the appropriate low-energy
effective theory remain unanswered. Since the phase is insulating, an effective model where charge fluctuations are treated as virtual excitations should be possible. One step forward in this direction has been taken by Motrunich \cite{Motrunich05},
who proposed to describe the SL phase with 4-spin interactions. However, whether a description in terms of a pure spin model is possible is far from obvious, in particular since there seems to be a jump in the
double occupancy at the transition from the three-sublattice N\'eel phase to the SL \cite{Morita02,Yoshioka09}.

In this Letter, we show that the correct low-energy theory of both insulating phases, and in particular of the
SL phase, is indeed a pure spin model. This has been achieved by deriving an effective spin
model to very high order about the strong coupling limit using perturbative continuous unitary transformations (PCUTs),
and by showing that it gives a qualitative and quantitative account of the transition from the three-sublattice
magnetic order to the SL state. This description gives deep insight into the nature of the
SL phase and clearly provides the appropriate framework for further studies.

The starting point is the single-band Hubbard model on the triangular lattice defined by the Hamiltonian
\begin{eqnarray}
\mathcal{H} &=& H_U+H_t\nonumber\\
            &=& U\sum_{i}n_{i\uparrow}n_{i\downarrow} -t\sum_{\langle i,j\rangle,\sigma}(c_{i\sigma}^{\dagger}c_{j\sigma}+\text{h.c.})
\label{Hubbard_model}
\end{eqnarray}
where the sum over $i$ runs over the sites of a triangular lattice, and the sum over $\langle i,j\rangle$ over
pairs of nearest neighbors.
To derive an effective low-energy Hamiltonian, we use PCUTs~\cite{Stein97,Knetter00,Knetter03_1,Reischl04} about the localized limit treating the hopping term as a perturbation.
The kinetic part can be written as $H_t=t\left( T_{0}+T_{1}+T_{-1}\right)$ where $T_{m}$ changes the number of doubly-occupied sites by $m$. The PCUT method provides order by order in $t/U$ an effective Hamiltonian $H_{\rm eff}$ with the
property $[H_{\rm U},H_{\rm eff}]=0$, i.e. $H_{\rm eff}$ is block-diagonal in the number of doubly-occupied sites.
The low-energy physics at half filling is expected to be contained in the block with no doubly-occupied site $H_{\rm eff}^{0}$, which can be expressed as a spin 1/2 model of the form
\begin{eqnarray}
 &&H_{\rm eff}^{0} = \mathrm{const} + \sum_{\vec{r},\vec{n}} J_{\vec{n}} \left(\vec{S}_{\,\vec{r}} \cdot\vec{S}_{\,\vec{r}+\vec{n}}\right)\\
                &&+ \sum_{\vec{r},\vec{n}_1,\vec{n}_2,\vec{n}_3} J^{\vec{n}_3}_{\vec{n}_1,\vec{n}_2} \left( \vec{S}_{\,\vec{r}}\cdot\vec{S}_{\,\vec{r}+\vec{n}_1}\right) \left( \vec{S}_{\,\vec{r}+\vec{n}_2} \cdot\vec{S}_{\,\vec{r}+\vec{n}_3}\right)+ \ldots \nonumber
\end{eqnarray}
where $\vec{r}$ and $\vec{r}+\vec{n}_{\alpha}$ denote sites on the triangular lattice. All remaining terms can be written in a similar way as products of $\vec{S}_{\,\vec{r}}\cdot \vec{S}_{\,\vec{r}'}$ due to
the SU(2) symmetry of the Hubbard model.

The PCUT provides the magnetic exchange couplings as series expansions in $t/U$. Since the spectrum of the Hubbard model is symmetric at half filling under the exchange $t\leftrightarrow -t$, only even order contributions are present \cite{MacDonald88}. We have determined all 2-spin, 4-spin and 6-spin interactions up to order 12.
%Interactions involving more than six spins have been calculated up to order 10.
The obtained series are valid in the thermodynamic
limit. To this end we have fully exploited for the first time in a PCUT calculation the linked cluster theorem
by using graph theory.
At order 12 there are 1336 topologically different graphs. The major complication for the determination
of higher orders comes from the large number of different spin operators, which requires to calculate all possible matrix elements for each graph.

%
% Size of the relevant spin couplings as function of t/U
%
%Figure PCUT Extrapolations Relevant Couplings
%%%%%%%%%%%%%%%%%%%%%%%%%%%%%%%%%%%%%%%%%%%%%%
\begin{figure}
\begin{center}
\includegraphics*[width=0.94\columnwidth]{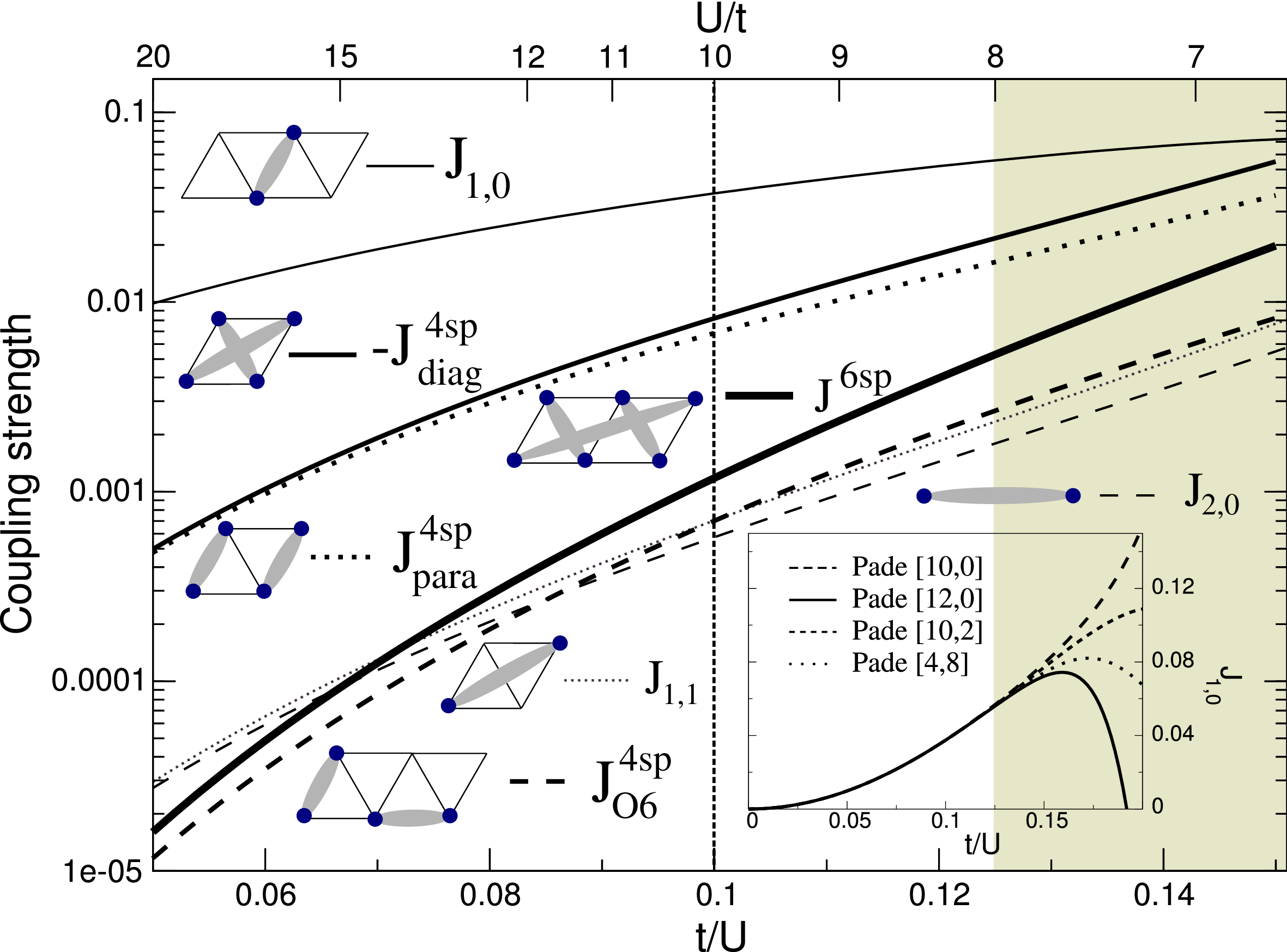}
\end{center}
\caption{(Color online) The most relevant exchange couplings displayed as a function of $t/U$. The 12th order bare series of all couplings appearing up to order 4 and the largest 4-spin interaction $J^{\rm 4sp}_{\rm O6}$ plus the largest 6-spin interaction $J^{\rm 6sp}$ appearing in order 6 are plotted. Grey shaded area marks the region where extrapolation effects become important. Vertical dashed line indicates the magnetic phase transition. Inset: various Pad\'e approximants as a function of $t/U$ for the nearast-neighbor Heisenberg exchange $J_{1,0}$.}
\label{fig:couplings}
\end{figure}
%%%%%%%%%%%%%%%%%%%%%%%%%%%%%%%%%%%%%%%%%%%%%%
The most relevant spin couplings are shown in Fig.~\ref{fig:couplings} as a function of $t/U$. As already pointed out in Ref.~\onlinecite{Motrunich05}, the most important correction to the nearest-neighbor Heisenberg exchange at order 4 are the 4-spin interactions on a diamond plaquette. High-order contributions result in more complex multi-spin interactions with larger spatial extension but also lead to renormalizations of already existing couplings. The latter effect causes for example a sizable splitting of the two 4-spin interactions shown in Fig.~\ref{fig:couplings}, which have the same absolute value at order 4. Processes involving sites far apart have small amplitudes, and multi-spin interactions between sites not far from each other are clearly the dominant corrections to the spin operators at order 4.

At large $U/t$, nearest-neighbour exchange dominates, and the system
is expected to develop three-sublattice long-range order \cite{Bernu92,Capriotti99,White07}. Using exact diagonalizations (ED) of clusters up to
36 sites, we have investigated how the properties of the model evolve upon reducing $U/t$. The most striking
feature is a level crossing in the ground state (GS) at $U/t\simeq 10$ (see Fig.~\ref{fig:ergpersite}). This level crossing is observed for
all cluster sizes at essentially the same value, which indicates the presence of a first order transition
in the thermodynamic limit. Since this ratio is comparable to the critical ratio where other methods have detected a
transition into a SL phase, it seems plausible that this level crossing corresponds to this
transition. Further investigations below $U/t\simeq 10$ confirm this guess (see below), but
let us first investigate to which extent the effective spin model provides an accurate description in this
parameter range. The proof that this is so relies on four observations:
%
%%%%%%%%%%%%%%%%%%%%%%%%%%%%%%%%%%%%%%%%%%%%%%
\begin{figure}
\begin{center}
\includegraphics*[width=\columnwidth]{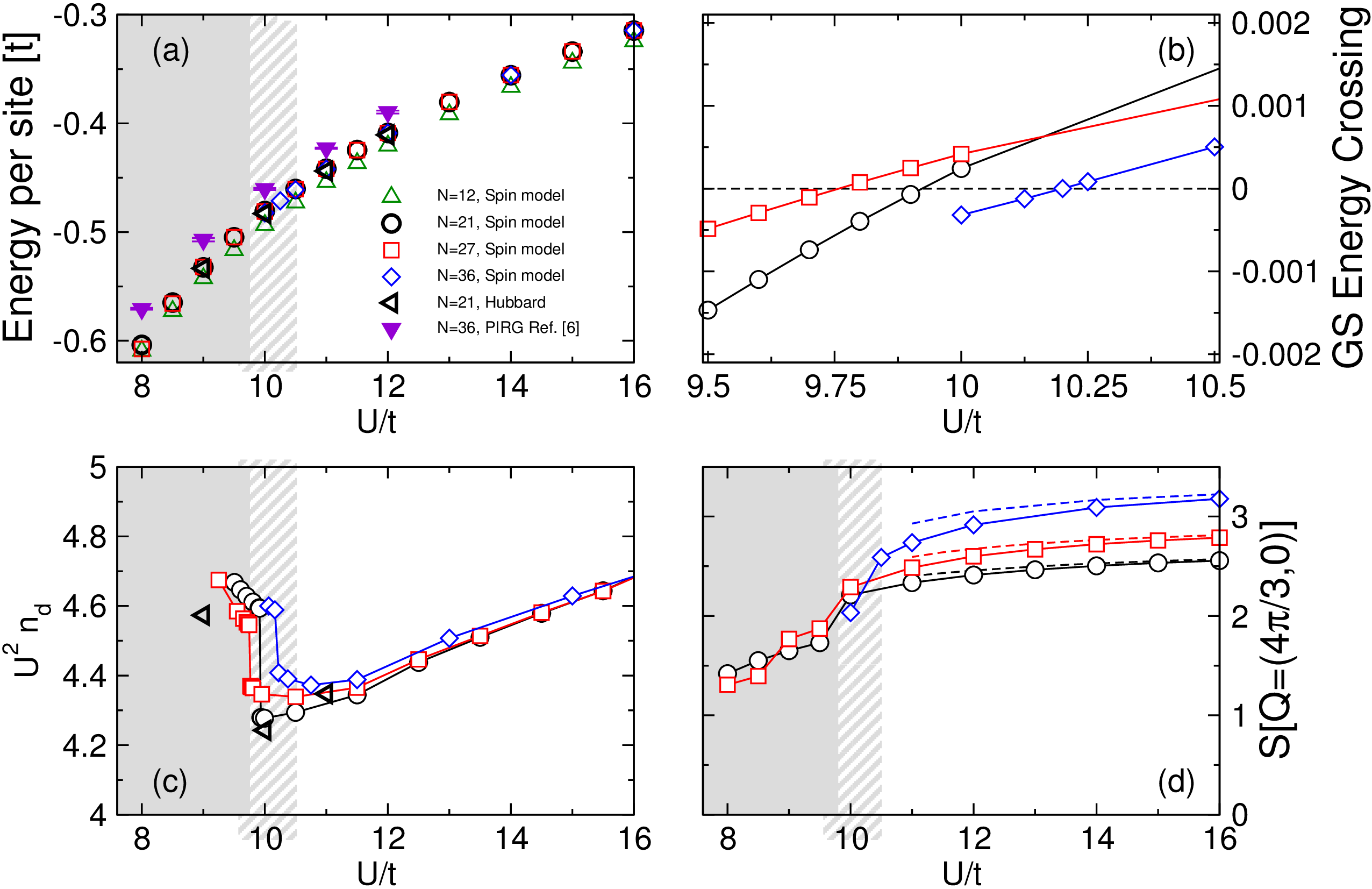}
\end{center}
\caption{(Color online) (a) Energy per site as a function of $U/t$ for the Hubbard model and the effective spin model. (b) Level crossings in the effective spin model signalled by the energy difference of the two lowest energies. (c) Double occupancy (times $U^2$) as a function of $U/t$.
(d) Magnetic structure factor at the 120$^\circ$ AFM ordering wave vector. Dashed lines denotes results for a simplified 2B-4B model evaluated up to order ten. Continuous lines with symbols are the results for the full 2B-4B-6B model considered here. The observable is up to order two in both cases.}
\label{fig:ergpersite}
\end{figure}
%%%%%%%%%%%%%%%%%%%%%%%%%%%%%%%%%%%%%%%%%%%%%%

- Convergence of the series: As usual, we use Pad\'e approximants to extrapolate the series, and we consider
that series are well converged as long as different Pade approximants give consistent results. Here this is definitely true up to $t/U\simeq 0.125$ ($U/t\simeq 8$), i.e. well beyond the level crossing
(see inset of Fig.~\ref{fig:couplings}). For all couplings, the absolute change between order 10 and 12 is below $10^{-4}$ ($10^{-3}$) for $t/U=0.1$ ($t/U=0.125$). In fact, clear indications of divergence only appear above $t/U\simeq 0.15$, and this is probably related to the metal-insulator transition.

- Comparison of GS energy with Hubbard model: In view of the anomaly observed at the transition by other approaches \cite{Morita02,Yoshioka09},
it is legitimate to ask whether the GS remains in the sector with no doubly-occupied site. To address this point, we have used ED of finite clusters to calculate the GS energy of both the Hubbard
model and the effective spin model as a function of $U/t$ (see Figs.~\ref{fig:ergpersite}a and \ref{fig:ergpersite}b) up to 21 sites for the Hubbard model, and up to 36 sites for the effective spin model. For the spin model, the total number of spin operators is enormous, and truncating the hierarchy of operators is necessary. We have included in our calculation all spin operators appearing up to order 6. Additionally, we have omitted for a given ratio $t/U$ all terms whose coupling was smaller than $10^{-6}$. When embedded on our largest 36 site cluster, this corresponds to $\sim$ 16'000 distinct terms in total.
The resulting total energies per site are displayed in Fig.~\ref{fig:ergpersite}. The energy per site
of the effective spin model depends very little on the size, and for the 21-site cluster,
the energies of the Hubbard model and of the effective spin model are in very good agreement on both sides of the transition. This last observation clearly establishes that the effective model remains accurate in the SL
phase. The GS energies for both
the Hubbard model and the effective model are actually significantly lower than recent estimates based on PIRG~\cite{Yoshioka09}, see Fig.~\ref{fig:ergpersite}a.

- Double occupancy: The comparison of the GS energies suggests that the GS is in the sector of the effective model with no doubly-occupied site in the $U/t$ range considered. This result seems to be incompatible with the jump of the double occupancy $n_d$ reported by other methods for the Hubbard model \cite{Morita02,Yoshioka09}, but in fact it is not because,
when calculating the expectation value of observables from the effective model, one must apply the same unitary transformation \cite{Knetter03_1,Delannoy05} which, in the present case, leads to a non-zero value of $n_d$.
In practice, it is simpler to use the Feynman-Hellmann theorem which allows to deduce it directly from the energy per site as $n_d=\partial_U \langle \mathcal{H} \rangle/N$. The results are depicted in Fig.~\ref{fig:ergpersite}c. A jump
is clearly present at the transition for all finite-size samples as a consequence of the level crossing of the effective model, and its magnitude is consistent with that calculated for the Hubbard model on the 21-site cluster
(see Fig.~\ref{fig:ergpersite}c).
The significant difference between 21 and 27 sites does not allow to perform a reliable finite-size scaling, but this does not affect the conclusion that the effective spin model captures the double occupancy jump at the transition.

%%%%%%%%%%%%%%%%%%%%%%%%%%%%%%%%%%%%%%%%%%%%%%
\begin{figure}
\begin{center}
\includegraphics*[width=0.88\columnwidth]{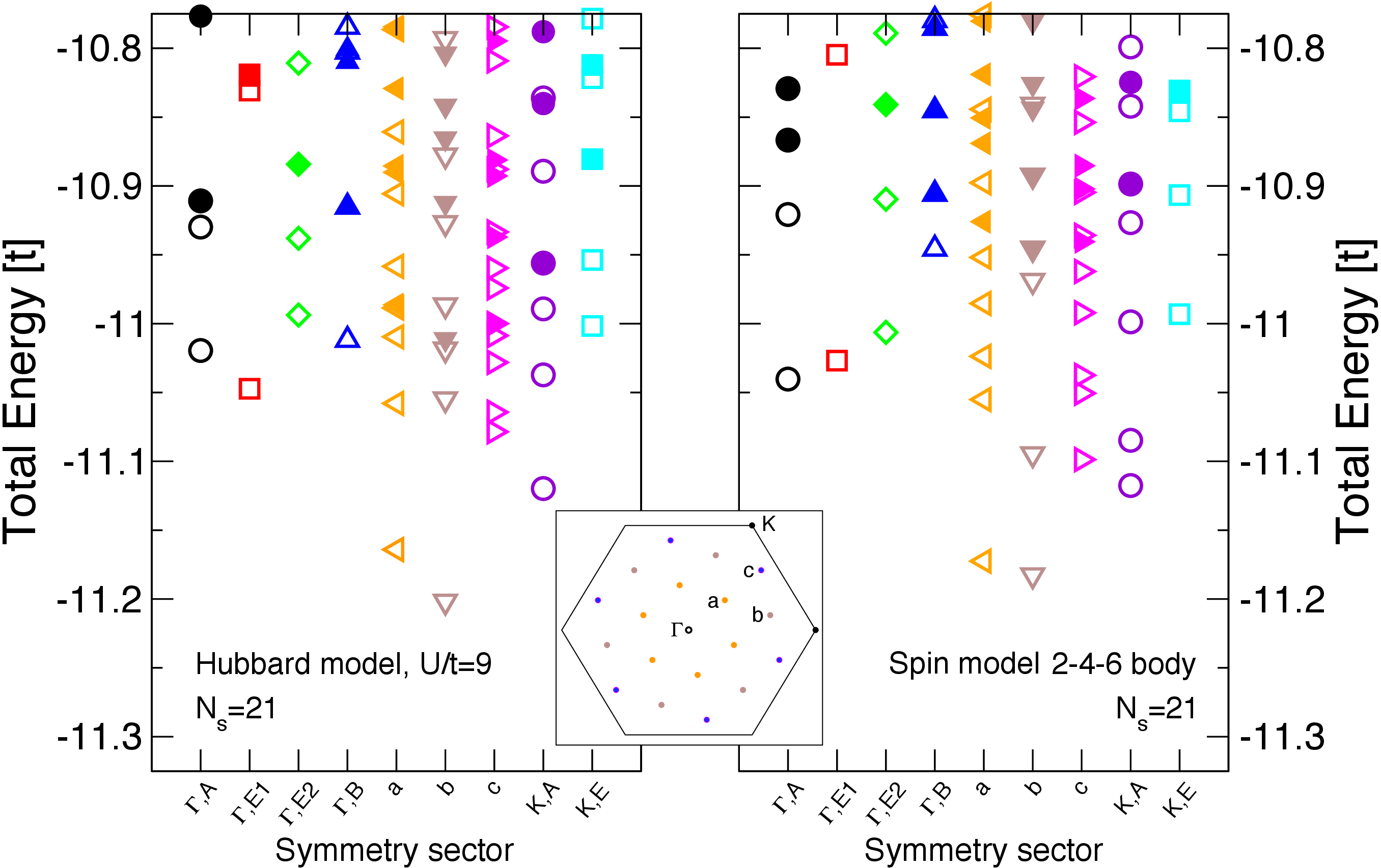}
\end{center}
\caption{(Color online) Comparison of the low-energy spectrum of a $N_s=21$ Hubbard model at $U/t=9$ (left panel)
to results obtained with the effective spin model (right panel). The $x$-axis labels the different symmetry sectors, while
empty (filled) symbols denote $S^z=1/2$ ($S^z=3/2$) levels. $a,b$ and $c$ label momenta in the interior of the Brillouin zone.}
\label{fig:lowenergyspectrum}
\end{figure}
%%%%%%%%%%%%%%%%%%%%%%%%%%%%%%%%%%%%%%%%%%%%%%

- Low-energy spectrum and GS momentum: It is also important to check how well the effective model reproduces the low-energy spectrum of the original model. To address this issue, we compare the low-energy spectrum at $U/t=9$ of a $N_s=21$ Hubbard model with the
low-energy excitations of the spin model on the same cluster~\footnote{The largest Hilbert space sector of the Hubbard
model (spin model) has $\dim\sim 6\times 10^9$ ($\dim\sim 10^5$).} in Fig.~\ref{fig:lowenergyspectrum}.
Qualitatively, the agreement is good. In particular, the non-trivial symmetry sectors of the GS
and of the first excited states are correctly reproduced.
Quantitatively, the difference between the energies (a fraction of a percent) is
dominated by the fact that the effective model on a finite cluster should
include, starting from order 6, a few processes which wrap around the cluster. These processes are
not taken into account in our effective model since they are not present in the thermodynamic limit.
The systematic error in the thermodynamic limit is much smaller since it only comes from the precision of
the coefficients and the neglect of truncated terms.

%%%%%%%%%%%%%%%%%%%%%%%%%%%%%%%%%%%%%%%%%%%%%%
\begin{figure}
\begin{center}
\includegraphics*[width=0.93\columnwidth]{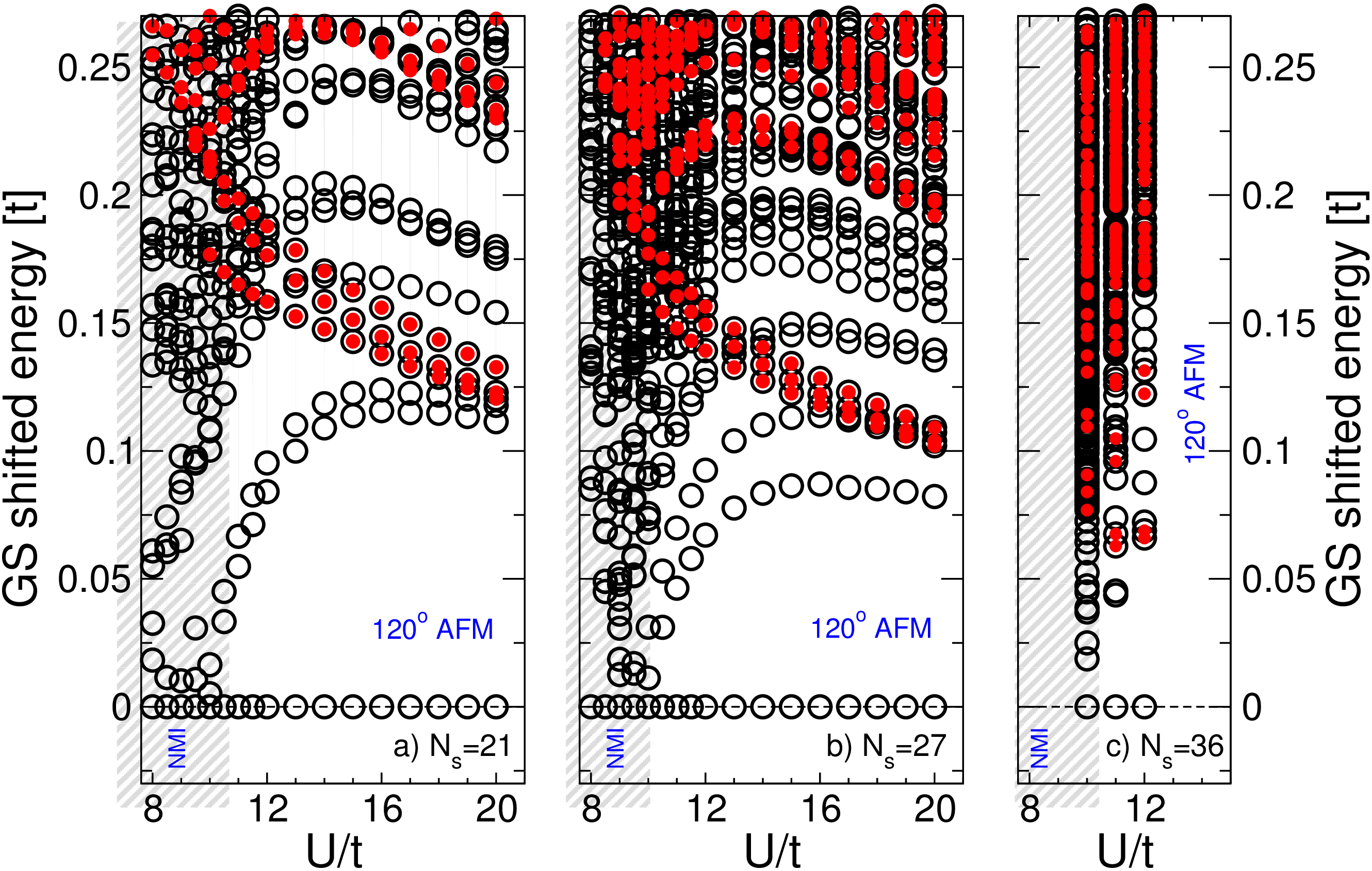}
\end{center}
\caption{(Color online) ED spectra of the truncated spin model for (a) $N_s=21$, (b) $N_s=27$, and (c) $N_s=36$ as a function of $U/t$. Open (black) symbols denote non-magnetic levels. Filled (red) symbols are energy levels corresponding to excitations with finite spin above the GS. The grey shaded area signals the extension of the SL phase.}
\label{fig:lowenergyspin}
\end{figure}
%%%%%%%%%%%%%%%%%%%%%%%%%%%%%%%%%%%%%%%%%%%%%%

The truncated spin model is thus an excellent starting point to investigate the quantum magnetism of the
Hubbard model on the triangular lattice. Consequently, we turn to a detailed discussion of its physical properties
as a function of $U/t$. The low-energy ED spectra for clusters of  $N_s=21, 27$, and $36$ sites are shown in Fig.~\ref{fig:lowenergyspin}.
For clarity the lowest level at each $U/t$ value is taken as the reference energy and set to zero, and the
spatial symmetry quantum numbers of the levels are not specified. For the 36-site cluster, the calculation of
the excitation spectrum is numerically very expensive, and it has been performed only for three values of $U/t$ close
to the transition, and for a simplified two- and four-spin coupling model (in contrast to the ground state energy
calculations in Fig.~\ref{fig:ergpersite}(a) which have been performed based on the more demanding model with the aforementioned cutoff $10^{-6}$). At large $U/t$ (down to $U/t\sim 15$), the level structure revealed by the 21 and 27-site clusters is typical of the tower of states of a 120$^\circ$ antiferromagnet (AFM)~\cite{Bernu92}. For smaller $U/t$ ratios, some levels start to bend down and to deviate from the leading $1/U$ behavior of the
Heisenberg regime. Drastic changes occur in the spectrum at $U/t\approx 10$. First of all, on {\it all} clusters,
excited states with different quantum numbers cross the large $U/t$ GS level. Besides, a large number of singlets accumulate at low energy in both the effective spin model {\it and} the Hubbard model for $U/t\leq 10$.

To better characterize this change of behaviour, we have calculated the
structure factor corresponding to the three-sublattice 120$^\circ$ order as a function of $t/U$ by implementing the unitary transformation on the observable at order $t^2/U^2$.
Above the level crossing, this structure factor
increases with the size, as expected for a long-range ordered phase. By contrast, below the level crossing
- i.e. in the putative SL region - the structure factor is finite size saturated, implying either short range correlations, or an algebraic decay with a sizeable decay exponent. At large $U/t$,
the slight reduction of the structure factor is mainly a
consequence of the admixture of empty and doubly occupied sites, which reduces the effective local spin length
$S^2$ below 3/4. However, as we approach the level crossing, the extra terms in the spin Hamiltonian further
reduce the structure factor, as can be seen by comparing the results for the 'full' effective model with those for a truncated model where only the 2-spin and 4-spin terms that appear at fourth order (but evaluated using
the full series) are kept (see Fig.\ref{fig:ergpersite}d). So one cannot exclude that the long-range magnetic order
disappears slightly before the level crossing.

The issue of the spin gap in the SL phase requires a careful discussion. Indeed, the energy of the first
magnetic excitation is nearly constant between 21 and 27 sites and drops dramatically for 36 sites. So depending
upon whether one includes the 36 cluster or not, finite-size effects suggest a vanishing gap or a finite gap.
Now, the significant difference between the 36-site cluster and the other two might be
due to an even-odd $N_s$ effect, or to the use of a simplified model, and it might be better not to include it in the finite-size analysis.
Now, concentrating on the 21 and 27 site clusters, the energy of the first magnetic excitation is indeed
almost constant, but if one looks carefully at the drastic $U/t$ dependence of the excited magnetic levels when approaching the SL regime from large $U/t$, some excitations bend down, and their energy decreases
very significantly between 21 and 27 sites, possibly leading to a gap closing in the thermodynamic limit.
%Generally speaking it is difficult to track excitation gaps above the respective ground states when a ground state level crossing occurs.
So it is difficult to decide on the presence of a spin gap based on the available ED data alone. In the next paragraph we show that the data can be interpreted
from a broader perspective.

In his seminal paper Motrunich~\cite{Motrunich05} proposed a plain Gutzwiller projected Fermi sea as a prototype spin wave function for the SL phase on the triangular lattice. This idea
later materialized into a theory of Spin Bose Metal (SBM) phases~\cite{Sheng09} (see Ref.~\cite{Lee05} for a related scenario), which has been successfully applied to a one-dimensional
triangular strip model with two- and four-spin interactions~\cite{Klironomos07,Sheng09}. Interestingly, the numerical low-energy spectrum of the triangular strip in the SBM phase looks
qualitatively similar to the spectrum we observe on the
triangular lattice, with respect to both the dense excitation spectrum with many low energy singlets and the irregular finite size behavior of the spin gap~\cite{Klironomos07}. A second remarkable
agreement between the SBM picture by Motrunich and our ED analysis comes from the quantum numbers of the ground state wave function in the SL region. On the $N_s=27$ site sample
for example the noninteracting half filled Fermi sea has a sixfold degenerate ground state momentum. This
prediction matches precisely the ground state quantum number of the effective spin model in the SL region adjacent to the magnetically ordered phase.
On the other samples ($N_s=21,36$) the Fermi sea ground state degeneracy is larger and thus less constraining, but the ground state quantum numbers of the SL regime are still compatible with this picture.

In conclusion, we have derived the appropriate low-energy theory of the non-magnetic phase of the
half-filled Hubbard model on the triangular lattice. It is a pure spin model with multi-spin exchange.
Furthermore, we have performed an ED investigation of this model. Several aspects of the results are compatible with the SBM picture predicting a gapless spectrum with a spinon Fermi surface. It will be interesting to extend our
investigation to the anisotropic triangular lattice \cite{Kandpal09,Nakamura09}, and to compare the results directly to the SL
material $\kappa$-(BEDT-TTF)$_{2}$Cu$_{2}$(CN)$_{3}$ and to the other members of the family. Finally, this approach is also applicable to other, possibly frustrated lattice topologies for which similarly interesting Mott phases can be expected.

\acknowledgments We thank R.~McKenzie for fruitful discussions and T.~Yoshioka for providing PIRG data for comparison.
K.P.S. acknowledges ESF and EuroHorcs for funding through his EURYI. The ED have been enabled
through computing time allocated at ZIH TU Dresden and the MPG RZ Garching.
F.M. acknowledges the financial support of the Swiss National Fund and of MaNEP.


\begin{thebibliography}{13}
\expandafter\ifx\csname natexlab\endcsname\relax\def\natexlab#1{#1}\fi
\expandafter\ifx\csname bibnamefont\endcsname\relax
  \def\bibnamefont#1{#1}\fi
\expandafter\ifx\csname bibfnamefont\endcsname\relax
  \def\bibfnamefont#1{#1}\fi
\expandafter\ifx\csname citenamefont\endcsname\relax
  \def\citenamefont#1{#1}\fi
\expandafter\ifx\csname url\endcsname\relax
  \def\url#1{\texttt{#1}}\fi
\expandafter\ifx\csname urlprefix\endcsname\relax\def\urlprefix{URL }\fi
\providecommand{\bibinfo}[2]{#2}
\providecommand{\eprint}[2][]{\url{#2}}

%\bibitem[{\citenamefont{Fazekas and P.~W}(1974)}]{Fazekas74}
%\bibinfo{author}{\bibfnamefont{P.}~\bibnamefont{Fazekas}} \bibnamefont{and}
%  \bibinfo{author}{\bibfnamefont{P.~W.}~\bibnamefont{Anderson}},
%  \bibinfo{journal}{Phil. Mag. A} \textbf{\bibinfo{volume}{30}},
%  \bibinfo{pages}{423} (\bibinfo{year}{1974}).

%\bibitem[{\citenamefont{Anderson}(1987)}]{Anderson87}
%\bibinfo{author}{\bibfnamefont{P.~W.} \bibnamefont{Anderson}},
%  \bibinfo{journal}{Science} \textbf{\bibinfo{volume}{235}},
%  \bibinfo{pages}{1196} (\bibinfo{year}{1987}).

%\bibitem[{\citenamefont{Georges, Kotliar, Krauth and Rozenberg}}]{Georges96}
%\bibinfo{author}{\bibfnamefont{A.}~\bibnamefont{Georges}},
%  \bibinfo{author}{\bibfnamefont{G.} \bibnamefont{Kotliar}},
%\bibinfo{author}{\bibfnamefont{W.} \bibnamefont{Krauth}},
%  \bibnamefont{and} \bibinfo{author}{\bibfnamefont{M.~J.} \bibnamefont{Rozenberg}},
%  \bibinfo{journal}{Rev. Mod. Phys.} \textbf{\bibinfo{volume}{68}},
%  \bibinfo{pages}{13} (\bibinfo{year}{1996}).

\bibitem{Georges96}  A. Georges {\it et al.}, Rev. Mod. Phys. {\bf 68}, 041101 (1996).

\bibitem{Morita02} H. Morita, S. Watanabe, and M. Imada, J. Phys. Soc. Jpn. {\bf 71}, 2109 (2002).

\bibitem[{\citenamefont{Kyung and Tremblay}(2006)}]{Kyung06}
\bibinfo{author}{\bibfnamefont{B.}~\bibnamefont{Kyung}} \bibnamefont{and}
  \bibinfo{author}{\bibfnamefont{A.M.S.} \bibnamefont{Tremblay}},
  \bibinfo{journal}{Phys. Rev. Lett.} \textbf{\bibinfo{volume}{97}},
  \bibinfo{pages}{046402} (\bibinfo{year}{2006}).

\bibitem[{\citenamefont{Sahebsara and S\'en\'echal}(2008)}]{Sahebsara08}
\bibinfo{author}{\bibfnamefont{P.}~\bibnamefont{Sahebsara}} \bibnamefont{and}
  \bibinfo{author}{\bibfnamefont{D.}~\bibnamefont{S\'en\'echal}},
  \bibinfo{journal}{Phys. Rev. Lett.} \textbf{\bibinfo{volume}{100}},
  \bibinfo{pages}{136402} (\bibinfo{year}{2008}).

%\bibitem{Tocchio08} L.F. Tocchio, F. Becca, A. Parola, and S. Sorella, Phys. Rev. B. {\bf 78}, 041101 (2008).
\bibitem{Tocchio08}  L.F. Tocchio {\it et al.}, Phys. Rev. B. {\bf 78}, 041101 (2008).

\bibitem[{\citenamefont{Yoshioka et~al.}(2009)\citenamefont{Yoshioka, Koga, and
  Kawakami}}]{Yoshioka09}
\bibinfo{author}{\bibfnamefont{T.}~\bibnamefont{Yoshioka}},
  \bibinfo{author}{\bibfnamefont{A.}~\bibnamefont{Koga}}, \bibnamefont{and}
  \bibinfo{author}{\bibfnamefont{N.}~\bibnamefont{Kawakami}},
  \bibinfo{journal}{Phys. Rev. Lett.} \textbf{\bibinfo{volume}{103}},
  \bibinfo{pages}{036401} (\bibinfo{year}{2009}).

%\bibitem{Meng10}  Z.Y. Meng, T. C. Lang, S. Wessel, F. F. Assaad, and A. Muramatsu, Nature {\bf 464}, 847 (2010).

\bibitem{Meng10}  Z.Y. Meng {\it et al.}, Nature {\bf 464}, 847 (2010).

%\bibitem{Shimizu03} Y.S. Shimizu, K. Miyagawa, K. Kanoda, M. Maesato, and G. Saito, Phys. Rev. Lett. {\bf 91}, 107001 (2003).
\bibitem{Shimizu03} Y.S. Shimizu {\it et al.}, Phys. Rev. Lett. {\bf 91}, 107001 (2003).

\bibitem[{\citenamefont{Motrunich}(2005)}]{Motrunich05}
\bibinfo{author}{\bibfnamefont{O.~L.} \bibnamefont{Motrunich}},
  \bibinfo{journal}{Phys. Rev. B} \textbf{\bibinfo{volume}{72}},
  \bibinfo{pages}{045105} (\bibinfo{year}{2005}).

%\bibitem{Misguich98} G. Misguich, B. Bernu, C. Lhuillier, and C. Waldtmann, Phys. Rev. Lett. {\bf 81}, 1098 (1998).

\bibitem[{\citenamefont{Stein}(1997)}]{Stein97}
\bibinfo{author}{\bibfnamefont{J.}~\bibnamefont{Stein}}, \bibinfo{journal}{J.
  Stat. Phys.} \textbf{\bibinfo{volume}{88}}, \bibinfo{pages}{487}
  (\bibinfo{year}{1997}).

\bibitem[{\citenamefont{Knetter and Uhrig}(2000)}]{Knetter00}
\bibinfo{author}{\bibfnamefont{C.}~\bibnamefont{Knetter}} \bibnamefont{and}
  \bibinfo{author}{\bibfnamefont{G.~S.} \bibnamefont{Uhrig}},
  \bibinfo{journal}{Eur. Phys. J. B} \textbf{\bibinfo{volume}{13}},
  \bibinfo{pages}{209} (\bibinfo{year}{2000}).

\bibitem[{\citenamefont{Knetter et~al.}(2003)\citenamefont{Knetter, Schmidt,
  and Uhrig}}]{Knetter03_1}
\bibinfo{author}{\bibfnamefont{C.}~\bibnamefont{Knetter}},
  \bibinfo{author}{\bibfnamefont{K.~P.} \bibnamefont{Schmidt}},
  \bibnamefont{and} \bibinfo{author}{\bibfnamefont{G.~S.} \bibnamefont{Uhrig}},
  \bibinfo{journal}{J. Phys. A} \textbf{\bibinfo{volume}{36}},
  \bibinfo{pages}{7889} (\bibinfo{year}{2003}).

\bibitem{Reischl04} A. Reischl, E. M\"uller-Hartmann, and G.S. Uhrig, Phys. Rev. B {\bf 70}, 245124 (2004).

\bibitem{MacDonald88} A.~H. MacDonald, S. M. Girvin, and D. Yoshioka, Phys. Rev. B {\bf 37}, 9753 (1988).	

\bibitem{Bernu92} B. Bernu, C. Lhuillier, and L. Pierre, Phys. Rev. Lett. {\bf 69}, 2590 (1992).
\bibitem{Capriotti99} L. Capriotti, A. E. Trumper, and S. Sorella, Phys. Rev. Lett. {\bf 82}, 3899 (1999).		
\bibitem{White07} S.~R. White and A. L. Chernyshev, Phys. Rev. Lett. {\bf 99}, 127004 (2007).

%\bibitem{Delannoy05} J.-Y. P. Delannoy, M. J. Gingras, P. C. Holdsworth, and A.-M. S. Tremblay, Phys. Rev. B {\bf 72}, 115114 (2005).
\bibitem{Delannoy05} J.-Y. P. Delannoy {\it et al.}, Phys. Rev. B {\bf 72}, 115114 (2005).

%\bibitem{LiMing00} W. LiMing, G. Misguich, P. Sindzingre, and C. Lhuillier, Phys. Rev. B {\bf 62}, 6272 (2000).	
\bibitem{LiMing00} W. LiMing {\it et al.}, Phys. Rev. B {\bf 62}, 6272 (2000).


\bibitem{Sheng09} D.N.~Sheng, O.I.~Motrunich, and M.P.A.~Fisher, Phys. Rev. B {\bf 79}, 205112 (2009).

\bibitem{Lee05}
%U(1) Gauge Theory of the Hubbard Model: Spin Liquid States and Possible Application to ?-(BEDT-TTF)2Cu2(CN)3
S.-S.~Lee and P.A.~Lee,
Phys. Rev. Lett. {\bf 95}, 036403 (2005).

\bibitem{Klironomos07} A. D. Klironomos {\it et al.},
%, J. S. Meyer, T. Hikihara, and K. A. Matveev,
Phys. Rev. B {\bf 76}, 075302 (2007).


%\bibitem[{\citenamefont{Kandal et~al.}(2009)\citenamefont{Kandal, Opahle,
%  Zhang, Jeschke, and Valenti}}]{Kandpal09}
%\bibinfo{author}{\bibfnamefont{H.~C.} \bibnamefont{Kandal}},
%  \bibinfo{author}{\bibfnamefont{I.}~\bibnamefont{Opahle}},
%  \bibinfo{author}{\bibfnamefont{Y.-Z.} \bibnamefont{Zhang}},
%  \bibinfo{author}{\bibfnamefont{H.~O.} \bibnamefont{Jeschke}},
%  \bibnamefont{and} \bibinfo{author}{\bibfnamefont{R.}~\bibnamefont{Valenti}},
%  \bibinfo{journal}{Phys. Rev. Lett.} \textbf{\bibinfo{volume}{103}},
%  \bibinfo{pages}{067004} (\bibinfo{year}{2009}).
\bibitem{Kandpal09} H.C. Kandal {\it et al.}, Phys. Rev. Lett. {\bf 103}, 067004 (2009).

%\bibitem{Nakamura09} K. Nakamura, Y. Yoshimoto, T. Kasugi, R. Arita, and M. Imada, J. Phys. Soc. Jap. {\bf 78}, 083710 (2009)
\bibitem{Nakamura09} K. Nakamura {\it et al.}, J. Phys. Soc. Jap. {\bf 78}, 083710 (2009)

\end{thebibliography}
\end{document}